\pdfoutput=1
\documentclass[trackchanges, twocolumn]{aastex7}
\usepackage[caption=false]{subfig}
\usepackage{graphicx}
\usepackage{bm}
\usepackage{amssymb,amsmath}
\usepackage{mathrsfs}
\usepackage{latexsym}
\usepackage{color}
\usepackage[normalem]{ulem}
\usepackage{dcolumn}
\usepackage[utf8]{inputenc}
\usepackage{microtype}
\usepackage{etoolbox}
\usepackage{accents}
\usepackage{multirow}
\usepackage{orcidlink}
\usepackage{makecell,tabularx,colortbl}
\usepackage{booktabs,cellspace}
\usepackage{multirow}
\usepackage{float}
\usepackage{xspace}

\definecolor{dodgerblue}{HTML}{1E90FF}
\definecolor{amethyst}{HTML}{a45ee5}
\definecolor{viennared}{HTML}{DA0A14}
\definecolor{ctorange}{HTML}{FF6C0C}
\definecolor{wales}{HTML}{ff0038}
\definecolor{benettongreen}{HTML}{009421}
\definecolor{ferrarired}{HTML}{ff2800}
\definecolor{austriawienpurple}{HTML}{441678}
\definecolor{gray}{HTML}{F0F0F0}
\definecolor{LightCyan}{rgb}{0.88,1,1}
\definecolor{headerblue}{RGB}{0,112,192}
\definecolor{lightgray}{gray}{0.9}
\definecolor{ZurichBlue}{rgb}{.255,.41,.884} 		
\definecolor{ZurichRed}{rgb}{0.9, 0.1, 0} 			
\definecolor{ZurichGreen}{rgb}{.196,.504,.396} 		
\definecolor{ZurichYellow}{rgb}{1,.648,0} 			
\definecolor{dodgerblue}{rgb}{0.12, 0.56, 1.0}
\definecolor{azure}{rgb}{0.0, 0.5, 1.0}
\definecolor{alizarincrimson}{rgb}{0.82, 0.1, 0.26}
\definecolor{mediumpurple}{rgb}{0.58, 0.44, 0.86}
\definecolor{lasallegreen}{rgb}{0.03, 0.47, 0.19}
\definecolor{my_gray}{rgb}{0,0,0}
\definecolor{uniwienblue}{HTML}{006699}
\definecolor{walesred}{HTML}{ff0038}
\definecolor{myorange}{HTML}{FF6C0C}
\definecolor{ferngreen}{HTML}{009246}
\definecolor{scarletred}{HTML}{CD212A}

\newcolumntype{a}{>{\columncolor{gray}}c}
\newcolumntype{b}{>{\columncolor{white}}c}

\newcommand{\massprimarymed}{11.5}
\newcommand{\massprimaryupper}{0.6}
\newcommand{\massprimarylower}{1.6}
\newcommand{\massprimary}{\massprimarymed^{+\massprimaryupper}_{-\massprimarylower}}

\newcommand{\masssecondarymed}{1.5}
\newcommand{\masssecondaryupper}{0.2}
\newcommand{\masssecondarylower}{0.1}
\newcommand{\masssecondary}{\masssecondarymed^{+\masssecondaryupper}_{-\masssecondarylower}}

\newcommand{\massratiomed}{0.13}
\newcommand{\massratioupper}{0.04}
\newcommand{\massratiolower}{0.01}
\newcommand{\massratio}{\massratiomed^{+\massratioupper}_{-\massratiolower}}

\newcommand{\totalmassmed}{13.1}
\newcommand{\totalmassupper}{0.6}
\newcommand{\totalmasslower}{1.4}
\newcommand{\totalmass}{\totalmassmed^{+\totalmassupper}_{-\totalmasslower}}

\newcommand{\chirpmed}{3.33}
\newcommand{\chirpupper}{0.09}
\newcommand{\chirplower}{0.07}
\newcommand{\chirpmass}{\chirpmed^{+\chirpupper}_{-\chirplower}}

\newcommand{\eccentricitymed}{0.145}
\newcommand{\eccentricityupper}{0.007}
\newcommand{\eccentricitylower}{0.063}
\newcommand{\eccentricity}{\eccentricitymed^{+\eccentricityupper}_{-\eccentricitylower}}

\newcommand{\meananomalymed}{3.2}
\newcommand{\meananomalyupper}{2.7}
\newcommand{\meananomalylower}{2.9}
\newcommand{\meananomaly}{\meananomalymed^{+\meananomalyupper}_{-\meananomalylower}}

\newcommand{\semimajormed}{757}
\newcommand{\semimajorupper}{8}
\newcommand{\semimajorlower}{26}
\newcommand{\semimajor}{\semimajormed^{+\semimajorupper}_{-\semimajorlower}}

\newcommand{\periastronmed}{270}
\newcommand{\periastronupper}{5}
\newcommand{\periastronlower}{20}
\newcommand{\periastron}{\periastronmed^{+\periastronupper}_{-\periastronlower}}

\newcommand{\chiprecmed}{0.06}
\newcommand{\chiprecupper}{0.13}
\newcommand{\chipreclower}{0.04}
\newcommand{\chiprec}{\chiprecmed^{+\chiprecupper}_{-\chipreclower}}

\newcommand{\chieffecmed}{0.005}
\newcommand{\chieffecupper}{0.071}
\newcommand{\chieffeclower}{0.095}
\newcommand{\chieffec}{\chieffecmed^{+\chieffecupper}_{-\chieffeclower}}

\newcommand{\spinprimarymed}{0.06}
\newcommand{\spinprimaryupper}{0.18}
\newcommand{\spinprimarylower}{0.05}
\newcommand{\spinprimary}{\spinprimarymed^{+\spinprimaryupper}_{-\spinprimarylower}}

\newcommand{\luminositydistancemed}{292}
\newcommand{\luminositydistanceupper}{107}
\newcommand{\luminositydistancelower}{122}
\newcommand{\luminositydistance}{\luminositydistancemed^{+\luminositydistanceupper}_{-\luminositydistancelower}}

\newcommand{\redshiftmed}{0.06}
\newcommand{\redshiftupper}{0.02}
\newcommand{\redshiftlower}{0.03}
\newcommand{\redshift}{\redshiftmed^{+\redshiftupper}_{-\redshiftlower}}

\newcommand{\eccentricityflowtwofivemed}{0.144}
\newcommand{\eccentricityflowtwofiveupper}{0.007}
\newcommand{\eccentricityflowtwofivelower}{0.056}
\newcommand{\eccentricityflowtwofive}{\eccentricityflowtwofivemed^{+\eccentricityflowtwofiveupper}_{-\eccentricityflowtwofivelower}}

\newcommand{\beq}{\begin{equation}}
\newcommand{\eeq}{\end{equation}}

\newcommand{\gw}{GW200105} 

\newcommand{\EFPE}{\textsc{pyEFPE}}
\newcommand{\XP}{\textsc{IMRPhenomXP}}
\newcommand{\XPHM}{\textsc{IMRPhenomXPHM}}

\newcommand{\uam}{Instituto de F\'isica Te\'orica UAM/CSIC, Universidad Aut\'onoma de Madrid, Cantoblanco 28049 Madrid, Spain}
\newcommand{\bham}{School of Physics and Astronomy and Institute for Gravitational Wave Astronomy, University of Birmingham, Edgbaston, Birmingham, B15 2TT, United Kingdom}
\newcommand{\aei}{Max Planck Institute for Gravitational Physics (Albert Einstein Institute), D-14467 Potsdam, Germany}

\begin{document}

\title{Orbital eccentricity in a neutron star -- black hole merger}

\author[orcid=0000-0002-9977-8546]{Gonzalo Morras}
\affiliation{\bham}
\affiliation{\uam}
\affiliation{\aei}
\email{gonzalo.morras@uam.es}

\author[orcid=0000-0003-4984-0775]{Geraint Pratten}
\affiliation{\bham}
\email{G.Pratten@bham.ac.uk}

\author[0000-0003-1542-1791]{Patricia Schmidt}
\affiliation{\bham}
\email[show]{P.Schmidt@bham.ac.uk}

\begin{abstract}
The observation of gravitational waves from merging black holes and neutron stars provides a unique opportunity to discern information about their astrophysical environment.
Two signatures that are considered powerful tracers to distinguish between different binary formation channels are general-relativistic spin-induced orbital precession and orbital eccentricity. 
Both effects leave characteristic imprints in the gravitational-wave signal that can be extracted from observations. 
To date, neither precession nor eccentricity have been confidently discerned in merging neutron star -- black hole binaries.
Here we report the measurement of orbital eccentricity in a neutron star -- black hole merger. 
Using, for the first time, a waveform model that incorporates precession and eccentricity, we perform Bayesian inference on the gravitational-wave event \gw{}~\citep{LIGOScientific:2021qlt} and infer a median orbital eccentricity of $e_{20}\sim \eccentricitymed$ at an orbital period of $0.1$s, ruling out eccentricities smaller than $0.028$ with $99.5$\% confidence. We find inconclusive evidence for the presence of precession, consistent with previous, non-eccentric results, but a more unequal mass ratio.
Our result implies a fraction of these binaries will exhibit orbital eccentricity even at small separations, suggesting formation through mechanisms involving dynamical interactions beyond isolated binary evolution.
Future observations will reveal the contribution of eccentric neutron star -- black hole binaries to the total merger rate across cosmic time.
\end{abstract}


\section{Introduction}
To date, the Laser Interferometer Gravitational-Wave Observatory (LIGO)~\citep{LIGOScientific:2014pky}, Virgo~\citep{VIRGO:2014yos} and KAGRA~\citep{KAGRA:2020tym} experiments have reported the detection of two binary neutron stars (BNS)~\citep{LIGOScientific:2017vwq, LIGOScientific:2020aai} and several neutron star--black hole (NSBH) binaries~\citep{LIGOScientific:2021qlt,KAGRA:2021vkt, LIGOScientific:2024elc,LIGOScientific:2025slb}. Here, we report the observation of orbital eccentricity in one of the NSBH binaries identified by LIGO and Virgo, GW200105\textunderscore{162426}, henceforth referred to as \gw.
A key feature of our analysis is the use of a novel theoretical waveform model that simultaneously incorporates both spin precession and orbital eccentricity~\citep{Morras:2025nlp}, thereby enabling the first joint measurements of both effects in NSBH binaries. 
As the complex interplay between spin-precession and orbital eccentricity can act as a source of systematic error~\citep{Romero-Shaw:2022fbf,Fumagalli:2023hde}, it is paramount that both effects are included in the analysis for an accurate measurement of the source properties.  

Orbital eccentricity and spin-precession encode vital information on the underlying astrophysical formation and evolution of these binaries. 
Understanding these evolutionary pathways is one of the most pressing challenges in the field, offering critical insight into stellar physics, binary formation mechanisms, environmental dynamics, and relativistic astrophysics. 
The standard formation scenario assumed for NSBH binaries is through the isolated evolution of interacting massive binary stars~\citep{Belczynski:2001uc,Broekgaarden:2021hlu,Broekgaarden:2021iew,Drozda:2020kqz,Xing:2023zfo}. 
Measuring large spin-orbit misalignment alone is insufficient to challenge this scenario, for example large natal kicks can source spin misalignment~\citep{Fragione:2021qtg, Baibhav:2024rkn}.
Orbital eccentricity, on the other hand, is inconsistent with isolated evolution which predicts that the binaries should rapidly circularize long before they enter the sensitivity band of ground-based GW detectors~\citep{Belczynski:2001uc,Broekgaarden:2021hlu,Broekgaarden:2021iew,Drozda:2020kqz,Xing:2023zfo}. 
As such, the orbital eccentricity observed in \gw{} implies that this event must have been formed through other formation channels involving dynamical interactions. 
In addition to the exceptional measurement of a non-zero eccentricity, our analysis also finds a more unequal mass ratio, bringing the mass of the secondary companion into better agreement with neutron star masses observed in the galaxy.

In this Letter, we present the analysis of \gw{} with a precessing-eccentric waveform model and argue that the observation of orbital eccentricity provides compelling evidence for dynamical formation scenarios. 
We further demonstrate the robustness of our eccentricity measurement through comprehensive systematic checks. 
In Section~\ref{sec:astro}, we outline some of the key astrophysical implications of this observation and how future detections of NSBH binaries with both ground- and space-based detectors will enable a robust characterisation of the astrophysical properties of the NSBH population. 
The importance of developing the models and tools to explore eccentric and precessing systems will only grow as next-generation GW detectors come online and significantly expand the observable population of such binaries.

\section{Event Selection and Data}
\label{sec:eventdata}
We analyse the two BNS events, GW170817~\citep{LIGOScientific:2017vwq} and GW190425~\citep{LIGOScientific:2020aai}, as well as the NSBH events reported in GWTC-3: GW200105, GW200115~\citep{LIGOScientific:2021qlt}, GW190426, GW190917~\citep{KAGRA:2021vkt}, and GW2305029~\citep{LIGOScientific:2024elc} from the fourth LIGO-Virgo-KAGRA (LVK) observing run. 
We do not analyse GW230518, which was observed during an engineering run~\cite{LIGOScientific:2025slb}.
Due to their low-mass nature, these events are inspiral-dominated, making them suitable for analysis with the \EFPE{} eccentric-precessing post-Newtonian (PN) waveform model~\citep{Morras:2025nlp}. 
We apply the Bayesian inference framework detailed below to these events using the publicly available strain data~\citep{LIGOScientific:2019lzm, KAGRA:2023pio}. 
Here, we focus on \gw{} due to its exceptional nature. The details for the other events can be found in~\citep{Morras:2025inprep}. 

The GW event \gw~\citep{GWdata,KAGRA:2023pio} was detected during the second half of the third observing run of Advanced LIGO and Virgo~\citep{aLIGO:2020wna, Virgo:2022ysc} on January 5, 2020 at 16:24:26 UTC. The noise power spectral density (PSD) for the LIGO Livingston and Virgo detectors at the time of the event were obtained from the same public data release and correspond to the ones used in the published LVK analyses~\citep{LIGOScientific:2021qlt}. The frequency range $46$ -- $51$ Hz is suppressed from the Virgo strain data via the PSD due to additional calibration systematics as per the original LVK analyses. 
LIGO Hanford was not observing at the time of the event.
The public strain data for LIGO Livingston underwent a data cleaning procedure~\citep{Cornish:2020dwh} to remove noise artifacts from scattered light below $25$ Hz. 
\gw{} was identified through multiple independent matched-filter searches with consistent signal-to-noise ratio (SNR) between the pipelines during the third observing run.
Due to the low SNR in Virgo, \gw{} was classified as a single-detector event, but was shown to clearly stand above background noise, shown in Fig. 3 of~\citep{LIGOScientific:2021qlt}.
Since its first detection, methodological advances in the estimation of the significance of GW events have placed the astrophysical origin of \gw{} on an even firmer footing, reporting a false alarm rate of $< 1.9 \times 10^{-5}$ per year~\citep{Davies:2022thw}. 
This is consistent with the independent analysis by~\citep{Nitz:2021zwj}, which also finds that the signal is clearly distinct from the noise background, building on the methodology developed in~\citep{Nitz:2020naa}.

\section{Methods}
\label{sec:methods}

\subsection{Waveform Models}
\label{app:waveform_models}
We determine the source properties of \gw{} by comparing the data against theoretical predictions of GWs from coalescing compact binaries.
To gauge potential systematic uncertainties and validate the robustness of our results, we analyse \gw{} with several waveform models that contain different physical effects: 
i) \textsc{TaylorF2Ecc}~\citep{Moore:2016qxz} -- a post-Newtonian inspiral-only waveform model for aligned-spin binaries on eccentric orbits;
ii) \textsc{pyEFPE}~\citep{Morras:2025nlp} -- a novel post-Newtonian inspiral-only waveform model that simultaneously takes into account spin-induced orbital precession and orbital eccentricity. 
The model excludes higher-order spherical harmonic modes (HOMs) of the radiation but it contains higher-order Fourier harmonics.
iii) \textsc{IMRPhenomXP}~\citep{Pratten:2020fqn, Pratten:2020ceb} -- a precessing inspiral-merger-ringdown (IMR) waveform model that describes the dominant harmonic mode of binaries on quasi-circular orbits; 
iv) \textsc{IMRPhenomXPHM}~\citep{Pratten:2020ceb, Garcia-Quiros:2020qpx} -- the same as iii) but including HOMs. 
In all three precessing cases, the orbital precession is modelled using the same \emph{effective} prescription combined with a multi-scale analysis (MSA)~\citep{Klein:2013qda}. 
The orbital eccentricity parameter used in \EFPE{} and quoted throughout this work is the post-Newtonian time-eccentricity as defined within the quasi-Keplerian parametrisation~\citep{Damour:1985}. 
Although alternative definitions of eccentricity are found in the literature, the one we report here is consistent with those commonly used in population studies~\citep{Vijaykumar:2024piy}.
For the models \textsc{TaylorF2Ecc}, \XP{} and \XPHM{}, we use their respective implementations in the public LVK Algorithm Library~\citep{lalsuite}. As \EFPE{} is the only model that includes precession and eccentricity, it represents our default choice.

\subsection{Parameter Estimation}
\label{app:parameter_estimation}
We perform coherent Bayesian inference to determine the source properties and to perform model comparison. Sampling is performed using the nested sampling~\citep{Skilling:2006gxv} algorithm \textsc{Dynesty}~\citep{dynesty} as implemented in \textsc{Bilby}~\citep{bilby_paper, bilby_pipe_paper} to obtain the posterior probability distribution function (PDF) 
\begin{equation}
p(\bm{\lambda} \mid d) = \frac{\mathcal{L}(d \mid \bm{\lambda}) \pi(\bm{\lambda})}{\mathcal{Z}(d)} 
\end{equation}
where $\mathcal{L}(d \mid \bm{\lambda})$ denotes the likelihood of observing the GW data $d$ given the model parameters $\bm{\lambda}$, $\pi(\bm{\lambda})$ is the prior and $\mathcal{Z}(d)$ denotes the evidence. Assuming stationary Gaussian noise in each GW detector, the joint likelihood of observing a signal $h(\bm\lambda)$ in a detector network is given by (see e.g.~\citep{Veitch:2014wba})
\begin{equation}
    \mathcal{L}(d \mid \bm{\lambda}) \propto \exp{\Bigg( -\frac{1}{2} \sum_{i=1}^N  \langle h_i(\bm \lambda) - d_i \mid h_i(\bm \lambda) - d_i \rangle \Bigg)},
\end{equation}
where $N$ is the number of detectors and the angle bracket denotes the noise-weighted inner product, defined as
\begin{equation}
    \langle a \mid b \rangle = 4 \Re \int_{f_{\rm low}}^{f_{\rm high}} \frac{\tilde{a}(f) \tilde{b}^*(f)}{S_n(f)} df,
\end{equation}
where $\tilde{a}(f)$ and $\tilde{b}(f)$ are the Fourier transforms of the real-values functions $a(t)$ and $b(t)$, ${}^*$ denotes complex conjugation and $S_n(f)$ is the (one-sided) noise PSD of the detector. The likelihood integration is performed over the frequency interval $[f_{\rm low}, f_{\rm high}]$. 
Our default analyses use a high-frequency cutoff of $f_{\rm high}=340$ Hz corresponding approximately to the median frequency of the minimum energy circular orbit (MECO)~\citep{Pratten:2020fqn}
obtained from the publicly available high-spin \textsc{IMRPhenomXPHM} samples~\citep{GWdata}, while we set $f_{\rm low} = 20$ Hz unless stated otherwise. For some of the robustness checks we use the even more conservative lower MECO frequency bound of $280$ Hz. This high-frequency truncation of the likelihood integration is necessary as the two eccentric waveforms only model the inspiral portion of the GW signal. The original LVK analyses used IMR waveform models with $f_{\rm high}=4096$ Hz. We verified that our results are robust under the variation of $f_{\rm high}$, finding that additional high-frequency content primarily sharpens the posterior distribution without shifting its median value or mode (see Fig.~\ref{fig:fvariation} and discussion in Appendix~\ref{app:parameter_estimation}). Moreover, despite the reduced high frequency cut-off we find a median network signal-to-noise ratio (SNR) of $13.7$ in the eccentric-precessing analysis. 

\textsc{Bilby} is the standard inference software adopted by the LVK, and our default sampler settings are consistent with those used in published LVK analyses. We also marginalise over frequency-dependent calibration uncertainties provided in~\citep{calibration}.
 
We choose priors templated on GWTC-1~\citep{LIGOScientific:2018mvr} for all non-eccentric parameters, allowing the dimensionless spin magnitude on both companions to be as high as $0.8$. 
We measure the eccentricity at a GW frequency of 20 Hz, and adopt a uniform prior $e_{20} \in \mathcal{U}(0, 0.4)$ as the default. Alternative eccentricity prior choices investigated are log-uniform, inverse linear and inverse quadratic. The mean anomaly is sampled uniformly between $0$ and $2\pi$.

\section{Results}
\label{sec:results}
\begin{figure}[t!]
    \centering
    \includegraphics[width=1\columnwidth]{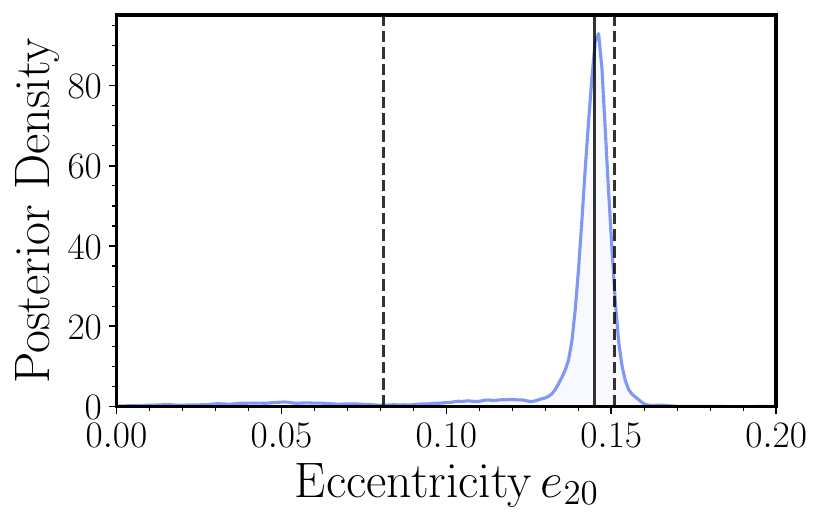}
    \caption{\textbf{Measured eccentricity distribution in \gw.} One-dimensional posterior probability distribution for our measurement of the orbital eccentricity at a GW frequency of $20$ Hz. The solid vertical line indicates the median and the two dashed lines the 90\% credible interval.
    }
    \label{fig:ecc}
\end{figure}

Using this inspiral-only eccentric-precessing waveform model~\citep{Morras:2025nlp}, we analysed the low-mass binary neutron star (BNS) and NSBH events~\citep{KAGRA:2021vkt, LIGOScientific:2017vwq, LIGOScientific:2020aai, LIGOScientific:2024elc} that are suitable for such an analysis. For \gw{} we find a median orbital eccentricity (with a 90\% credible interval) at a gravitational-wave (GW) frequency of $20$ Hz of $e_{20} = \eccentricity$.
The posterior distribution of our measurement of the orbital eccentricity is shown in Fig.~\ref{fig:ecc}. The lower limit of the 99\% highest posterior density interval is $0.028$, strongly disfavouring a circular binary.
However, we do not find statistically significant evidence for orbital eccentricity in any of the other analysed events, the details of which are reported elsewhere~\citep{Morras:2025inprep} and summarised in Fig.~\ref{fig:events} in Appendix~\ref{app:add_pe}. From the eccentricity measurement in \gw, we infer a median advance of periastron at a GW frequency of 20 Hz of $\dot{\omega} = \periastronmed\, {}^\circ{\rm s}^{-1}$, which is approximately eight orders of magnitude higher than the periastron advance reported for the eccentric double neutron star system PSR J1949+2052, $25.6{}^\circ {\rm yr}^{-1}$~\citep{Stovall:2018ouw}.
The measurement of non-zero orbital eccentricity in \gw{} provides compelling evidence that this NSBH formed through channels beyond standard isolated binary evolution. Plausible formation mechanisms for this binary include hierarchical (triple or quadruple) systems or dynamical interactions in dense stellar environments, representing the first such NSBH observation.
We determine that \gw{} has a median primary mass of $m_1 = \massprimarymed M_{\odot}$ and a median secondary mass of $m_2 = \masssecondarymed M_{\odot}$.
\begin{figure}[t!]
    \centering
    \includegraphics[width=1\columnwidth]{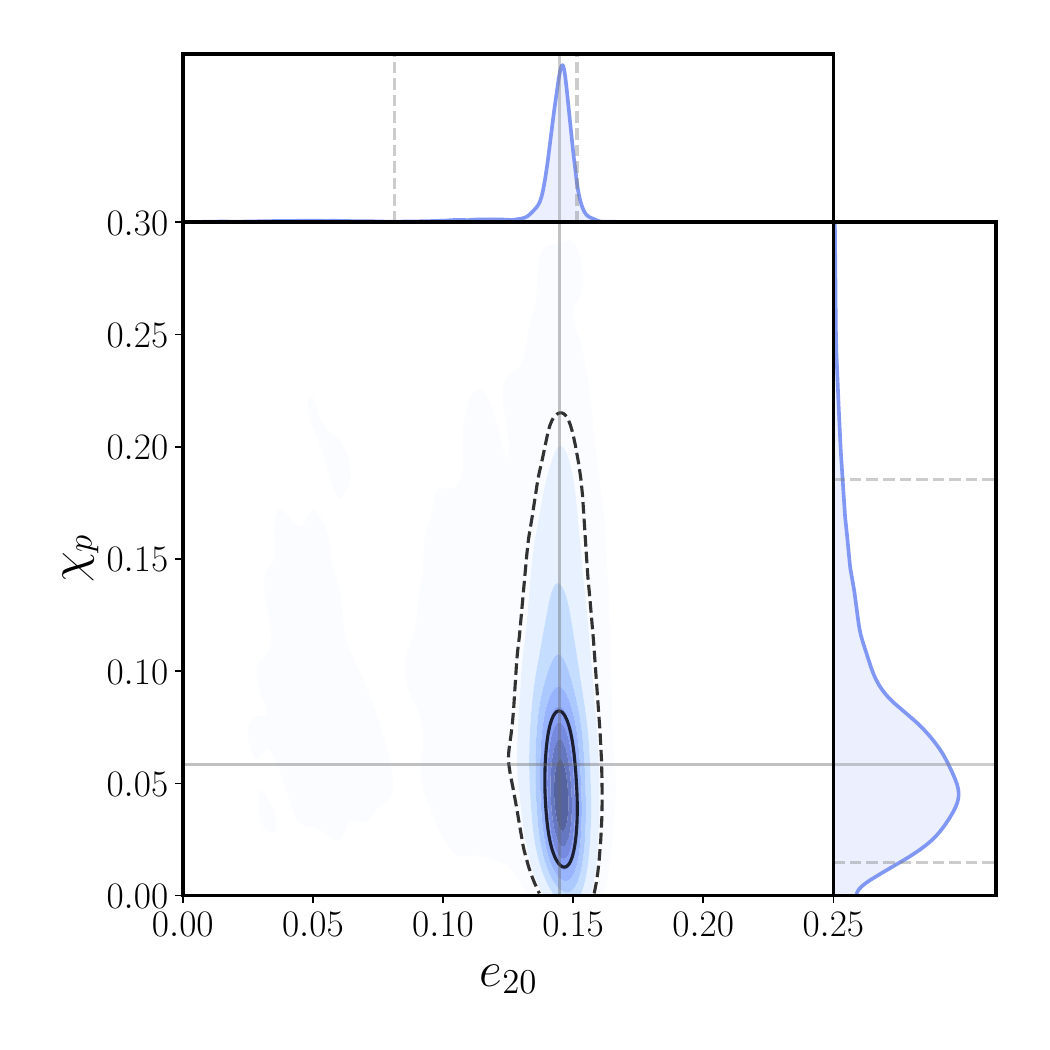}
    \caption{\textbf{Inferred eccentricity and precessing spin distributions in \gw.} One-dimensional and two-dimensional posterior probability distributions for our measurement of the orbital eccentricity and the precession spin $\chi_p$~\citep{Schmidt:2014iyl} at a GW frequency of $20$ Hz. The solid lines indicate the median, and the dashed lines the 90\% credible interval. We find no correlation between precession and eccentricity for \gw.
    }
    \label{fig:ecc-chip}
\end{figure}
The component masses we infer differ from those reported by the LVK collaboration~\citep{LIGOScientific:2021qlt}, who found a median black hole mass of $m_1 = 8.9 M_{\odot}$ and a neutron star mass of $m_2 = 1.9 M_{\odot}$.
Our result favours a lighter neutron star, consistent with observed millisecond pulsar masses~\citep{Antoniadis:2016hxz}, and a more massive black hole. 
This discrepancy aligns with the expected bias in mass parameters when eccentricity is neglected in GW analyses~\citep{Favata:2021vhw}, noting that our result is based on a PN model with limited information. 
During the inspiral, the gravitational-wave phase evolution is primarily determined by the chirp mass, $\mathcal{M}_c = (m_1 m_2)^{3/5} / (m_1 + m_2)^{1/5}$, making it one of the best constrained parameters. 
Neglecting eccentricity can bias the inferred chirp mass, such that $\mathcal{M}_c^{\rm biased} \approx \mathcal{M}_c (1 + 157 e^2_{20} / 40)$~\citep{Morras:2025nlp}, implying that the chirp mass will be over-estimated  if the system has non-zero eccentricity. 
This bias is consistent with the heavier chirp mass reported by the LVK, $\mathcal{M}_c = 3.41^{+0.08}_{-0.07} M_{\odot}$~\citep{LIGOScientific:2021qlt}, compared to the measurement presented here, $\mathcal{M}_c = \chirpmass M_{\odot}$.
Such systematic biases in the masses and eccentricity can propagate into tests of fundamental and nuclear physics with NSBH binaries~\citep{Sanger:2024axs}, highlighting the importance of including eccentricity in future analyses in light of this observation.

We find that the binary is consistent with very low, at most mildly precessing, spins with an effective spin along the orbital angular momentum~\citep{Ajith:2009bn} of $\chi_{\rm eff} = \chieffec$ at 90\% credibility and an effective precessing spin~\citep{Schmidt:2014iyl} of $\chi_p < 0.19$ at the $95$\% credible upper limit, consistent with the non-eccentric LVK analysis.
Our analysis allows us to simultaneously constrain precession and eccentricity. As shown in Fig.~\ref{fig:ecc-chip}, we find no correlation between eccentricity and precession in \gw{}.
The properties of \gw{} inferred from our analysis with the eccentric-precessing waveform model are listed in Table~\ref{tab:properties}; the one- and two-dimensional posterior distributions for mass and spin parameters and the eccentricity are shown in Fig.~\ref{fig:corner} in Appendix~\ref{app:add_pe}.

To assess the robustness of our non-zero orbital eccentricity measurement, we performed numerous checks pertaining to waveform uncertainties, parameter estimation uncertainties, prior choices and noise artifacts in the data. The details of these additional analyses and supporting investigations can be found in Appendix~\ref{app:robustness}.

We carried out a series of investigations to confirm that our non-zero eccentricity measurement is not a result of waveform systematics in \EFPE{}. These include checks on the model consistency in the non-precessing limit, the efficacy of the model in recovering eccentricity, and the impact of higher-order multipoles. All tests yield consistent outcomes and are detailed in Appendix~\ref{app:waveform_uncertainties}.
We further confirmed that the finite upper cut-off frequency in the likelihood integration due to the inspiral-only model does not change the eccentricity measurement nor do different sampler settings (see Appendix~\ref{app:pe_uncertainties}).

Our default analysis assumes a uniform prior on the eccentricity, which assigns equal weight to all eccentricities considered and allows the posteriors to be primarily driven by the data. 
We verified that the measurements are robust against alternative prior choices, except when adopting a highly informative prior, such as the log-uniform prior with a lower prior bound of $< 10^{-2}$, which has negligible support above $e \gtrsim 0.1$; see Appendix~\ref{app:prior_effects} for the details.
The log-uniform prior concentrates probability density at small eccentricities and strongly suppresses larger values. This makes posterior distributions particularly susceptible to the lower boundary condition, potentially allowing prior specifications to dominate likelihood information in low signal-to-noise regimes.
The impact of the prior choice can be reproduced when analysing mock theoretical signals consistent with the parameters inferred from \gw{}.  
We also explored alternative sampling strategies, including adopting a uniform prior on $e_{20}$ while sampling in $\log_{10}(e_{20})$, to ensure that a larger dynamical range of eccentricities is explored.
We find consistent agreement between the analyses, indicating that the sampling method does not affect the inferred posteriors.

By analysing non-eccentric injections in random realisations of Gaussian noise weighted by the event PSD, we did not find evidence that the measurement of non-zero orbital eccentricity in \gw{} is caused by random noise fluctuations (see Appendix~\ref{app:noise_artifacts}).
Using a hierarchical empirical Bayesian model for the noise-induced eccentricity, derived from theses simulations, we estimate the probability that noise alone could induce an eccentricity as large as \gw's median value to be $2.3 \times 10^{-4}$. This suggests that the eccentric result for \gw{} is highly unlikely to be caused by noise alone.

\section{Astrophysical Implications}
\label{sec:astro}
We now explore some of the astrophysical implications, arguing that the properties of the event are consistent with formation scenarios that involve dynamical interactions and standard stellar evolution processes that predict low black hole spins.
Understanding the formation and evolution of stellar-mass black holes and neutron stars remains one of the most pressing challenges in modern astrophysics.
While numerous evolutionary pathways have been proposed~\citep{Broekgaarden:2021iew}, the exact mechanisms that give rise to NSBH binaries are still not fully understood. 
Identifying a unique formation pathway would have profound implications for stellar physics and binary evolution.
Detecting these binaries in the electromagnetic (EM) spectrum remains challenging, with only a few possible candidates having been identified, for example~\citep{Barr:2024wwl}.
Gravitational-wave observations therefore afford us a critical opportunity to characterise astrophysical formation channels~\citep{LIGOScientific:2021qlt,LIGOScientific:2024elc}.

\begin{table}[t!]
\renewcommand{\arraystretch}{1.25}
\centering
    \begin{tabular}{l r}
         \hline
         \hline
         Primary mass, $m_1\, (M_\odot)$ & $\massprimary$ \\
         \hline
         Secondary mass, $m_2\, (M_\odot)$ & $\masssecondary$  \\
         \hline
         Total mass, $M=m_1+m_2\, (M_\odot)$ & $\totalmass$ \\
         \hline
         Mass ratio, $q = m_2/m_1$ & $\massratio$ \\
         \hline
         Chirp mass, $\mathcal{M}_c \, (M_\odot)$ & $\chirpmass$  \\
         \hline
         Primary spin, $a_1/m_1$ & $\spinprimary$ \\
         \hline
         Secondary spin, $a_2/m_2$ & undetermined \\
         \hline
         Effective inspiral spin, $\chi_{\rm eff}$ & $\chieffec$ \\
         \hline
         Effective precession spin, $\chi_p$ & $\chiprec$ \\
         \hline
         Orbital eccentricity, $e_{20}$ & $\eccentricity$ \\
         \hline
         Mean anomaly, $l\, (\rm rad)$ & $\meananomaly$ \\
         \hline
         Luminosity distance, $D_L\, (\rm Mpc)$ & $\luminositydistance$  \\
         \hline
         \hline
         Redshift, $z$ & $\redshift$  \\
         \hline
         Semi-major axis, $a$ (km) &  $\semimajor$\\
         \hline
         Rate of periastron advance, $\dot{\omega}\, ({}^\circ\, \rm{s}^{-1})$ & $\periastron$ \\
         \hline 
         \hline
    \end{tabular}
    \caption{\textbf{Properties of the eccentric NSBH binary \gw}. Measured and derived median source parameters and uncertainties at the 90\% credible interval from our analysis of \gw~ with the eccentric-precessing waveform model. The redshift, semi-major axis and the rate of periastron advance are derived quantities. All parameters are quoted at a GW frequency of $20$ Hz. 
    }
    \label{tab:properties}
\end{table}

Formation channels for NSBH binaries encompass a broad spectrum of mechanisms that can be broadly classified into two principal categories: isolated binary evolution and dynamical assembly. 
In isolated evolution, the binary evolves without external perturbations~\citep{Broekgaarden:2021iew}, whereas dynamical assembly typically occurs in dense stellar environments~\citep{Freire:2004sr,Fragione:2018yrb,Ye:2019luh,Rastello:2020sru,Trani:2021tan}. 
Higher-order multiplicity systems represent an intriguing intermediate case~\citep{Silsbee:2016djf,Toonen:2016htr,Antonini:2017ash,Fragione:2019zhm}, in which a tertiary companion can influence the binary’s evolution through secular gravitational interactions~\citep{Zeipel:1910,Lidov:1962,Kozai:1962}.
Unfortunately, component mass distributions can exhibit significant degeneracies across formation pathways~\citep{Broekgaarden:2021iew}, limiting their effectiveness in distinguishing between astrophysical formation channels.
In contrast, spin precession and orbital eccentricity serve as more robust tracers for the binary dynamics, predicting distinctive signatures that can be associated with dynamical formation processes~\citep{Fragione:2018yrb,Ye:2019luh,Rastello:2020sru,Trani:2021tan} or hierarchical triples (quadruples)~\citep{Silsbee:2016djf,Toonen:2016htr,Antonini:2017ash,Fragione:2019zhm}.

The currently favoured scenario for NSBH formation involves the isolated evolution of a massive binary~\citep{Belczynski:2001uc,Mapelli:2018wys,Drozda:2020kqz,Broekgaarden:2021iew,Mandel:2021ewy,Broekgaarden:2021hlu,Xing:2023zfo, vanSon:2024bxr}.
There are compelling arguments that black holes may form with low natal spins~\citep{Fuller:2019sxi}. In particular, angular momentum transport is expected to be efficient in redistributing momentum from the helium core to the hydrogen envelope~\citep{Fuller:2019sxi}, with subsequent envelope stripping effectively removing angular momentum from the progenitor.
Several mechanisms have been proposed that can give rise to non-negligible orbital eccentricity in isolated binaries. Moderate supernova kicks can induce natal orbital eccentricity, though excessively large kicks would disrupt the binary, significantly suppressing the merger rate~\citep{Wysocki:2017isg,Giacobbo:2019fmo}. 
Recent analyses have also argued that mass transfer can induce residual orbital eccentricity in a non-trivial fraction of binaries~\citep{Rocha:2024oqc}, challenging the common assumption that it circularises the binary.
Nonetheless, energy loss via gravitational radiation is extremely efficient at dissipating eccentricity at a rate $\propto (1 - e^2)^{-5/2}$~\citep{Peters:1963ux}, causing binaries to rapidly circularise. Consequently, orbital eccentricity is expected to be negligible by the time these systems enter the sensitive frequency band of the current ground-based GW detectors.
The canonical prediction in isolated stellar evolution is therefore for NSBH binaries with low spins and negligible eccentricity~\citep{Broekgaarden:2021iew}.
An important exception arises in hierarchical triples (or quadruples)~\citep{Silsbee:2016djf,Toonen:2016htr,Antonini:2017ash,Fragione:2019zhm}, in which the Zeipel–Lidov–Kozai mechanism~\citep{Zeipel:1910,Lidov:1962,Kozai:1962} can excite the binary towards highly eccentric orbits despite the GW emission. 

Dynamical interactions provide an alternative formation pathway for NSBH binaries, giving rise to a range of unique astrophysical phenomena.  
These include dynamical processes in globular clusters~\citep{Portegies:2000,Freire:2004sr}, formation in young star clusters~\citep{Rastello:2020sru,Santoliquido:2020bry,Trani:2021tan}, or formation in nuclear star clusters near galactic centers~\citep{Stephan:2016kwj,Fragione:2018yrb}. 
Current population models indicate that these dynamical channels may contribute less to the overall observed merger rate~\citep{LIGOScientific:2021qlt,Broekgaarden:2021iew}, especially in globular clusters~\citep{Phinney:1991apjl,Portegies:2000,Freire:2004sr,Ye:2019luh}. 
Formation in young star clusters and nuclear star clusters, however, predict rates that are more consistent with the isolated binary evolution channels whilst predicting non-trivial eccentricities~\citep{Rastello:2020sru,Trani:2021tan,Fragione:2018yrb}. 

One of the key predictions for dynamical formation channels is that spin orientations can be isotropic and that binaries can have non-negligible eccentricity at formation.
If, however, black holes are born with low natal spins, then these binaries could effectively be non-spinning and hence have negligible spin precession.
If the black hole is formed through repeated hierarchical mergers~\citep{Gerosa:2021mno}, the spins could be significantly larger, although this scenario is disfavoured by the small black hole spin measured in \gw{}.  

Due to the high efficiency at which gravitational radiation circularises binaries, the non-zero eccentricity measured in \gw{} requires that either a mechanism excites eccentricity throughout the inspiral or the binary is formed with large natal eccentricity at small orbital separations, such that there is insufficient time for eccentricity to be radiated away. 
This presents a challenge for conventional isolated evolution~\citep{Broekgaarden:2021iew}. 
While dynamical capture could explain the observed eccentricity~\citep{Samsing:2013kua}, such events are expected to be extremely rare. 
More compelling explanations include hierarchical triple interactions mediated by the Zeipel–Lidov–Kozai mechanism~\citep{Zeipel:1910,Lidov:1962,Kozai:1962}, which helps transfer angular momentum between the inner binary and the outer tertiary. 
This mechanism can sustain comparatively large eccentricities in a range of astrophysical environments, including hierarchical systems~\citep{Silsbee:2016djf,Toonen:2016htr,Antonini:2017ash,Fragione:2019zhm}, young stellar clusters~\citep{Trani:2021tan}, and nuclear star clusters, where a central supermassive black hole acts as the perturbing tertiary~\citep{Stephan:2016kwj,Fragione:2018yrb}.

Although a single observation cannot be used to determine the population properties, \gw{} provides direct evidence that NSBH binaries can form through dynamical interactions, and potentially at higher rates than previously anticipated. 
Moreover, observations of eccentric binaries can offer crucial constraints on the branching ratio between formation channels~\citep{Zevin:2021rtf}, eventually allowing us to disentangle the relative merger rates.

\begin{figure}[t!]
    \centering
    \includegraphics[width=1\columnwidth]{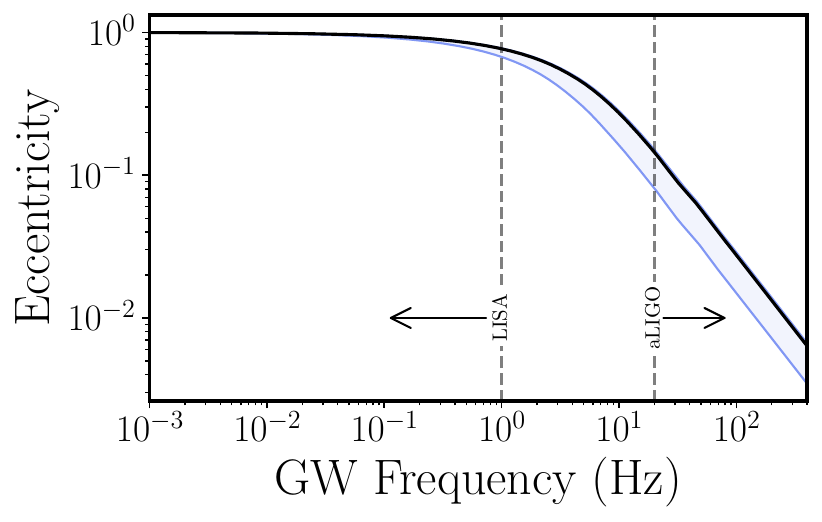}
    \caption{\textbf{Estimated eccentricity evolution of \gw.} The estimated orbital eccentricity of \gw~as a function of the GW frequency~\citep{Moore:2016qxz} between $1$ mHz and $400$ Hz and the corresponding orbital period in seconds (top axis), along with the $90\%$ credible interval obtained from our measurement, assuming that the evolution is driven entirely by GW emission. The arrows indicate the sensitive regions of Advanced LIGO ($f_{\rm GW} \geq 20$Hz) and LISA ($f_{\rm GW} \leq 1$Hz). 
    }
    \label{fig:ecc-evol}
\end{figure}

There are important implications for the NSBH binary population that we expect to be detectable in future gravitational wave observations, including the next-generation of ground-based detectors~\citep{Evans:2021gyd,Branchesi:2023mws} and at mHz frequencies in space-based observatories such as LISA~\citep{Klein:2022rbf,LISA:2024hlh}. 
Assuming the eccentricity evolution is driven entirely by gravitational radiation, for simplicity, then we observe that the eccentricity rapidly approaches unity towards the LISA band, as shown in Fig.~\ref{fig:ecc-evol}. 
This will be sensitive to exact astrophysical processes and interactions that couple to the eccentricity evolution.

\section{Conclusions}
Growing evidence for eccentricity in binary black hole observations~\citep{Romero-Shaw:2020aaj,Gamba:2021gap,Gayathri:2020coq,Romero-Shaw:2022xko,Gupte:2024jfe} has highlighted the importance of understanding the role of complex gravitational interactions in compact binary formation and evolution. 
The non-zero eccentricity measurement of $e_{20} = \eccentricity$ for the NSBH event \gw{} reported here provides new insights into the complex evolutionary pathways for NSBH systems, with significant implications for understanding compact object formation in both isolated and dynamical environments. 

The recent data release of GWTC-4.0~\citep{LIGOScientific:2025slb} provides limited opportunity to further explore the astrophysical properties of the NSBH population~\citep{LIGOScientific:2025pvj}.
Future gravitational-wave observations, however, will allow us to significantly improve our ability to distinguish between different formation channels, providing deeper insight into the evolutionary pathways for NSBH binaries and the complex astrophysical processes at work.

\section*{Acknowledgments}
The authors thank Alberto Vecchio and Michael Zevin for discussions and comments on the manuscript.
G.M. acknowledges support from the Ministerio de Universidades through Grants No. FPU20/02857 and No. EST24/00099, and from the Agencia Estatal de Investigaci\'on through the Grant IFT Centro de Excelencia Severo Ochoa No. CEX2020-001007-S, funded by MCIN/AEI/10.13039/501100011033.
G.M. acknowledges the hospitality of the University of Birmingham and the Institute for Gravitational Wave Astronomy. P.S. and G.P. are grateful to the Institute for Advanced Study for hospitality, where parts of this work were carried out.
G.P. is very grateful for support from a Royal Society University Research Fellowship URF{\textbackslash}R1{\textbackslash}221500 and RF{\textbackslash}ERE{\textbackslash}221015, and gratefully acknowledges support from an NVIDIA Academic Hardware Grant.
G.P. and P.S. acknowledge support from STFC grants ST/V005677/1 and ST/Y00423X/1, and a UK Space Agency grant ST/Y004922/1. 
P.S. also acknowledges support from a Royal Society Research Grant RG{\textbackslash}R1{\textbackslash}241327.
The authors are grateful for computational resources provided by the LIGO Laboratory (CIT) and supported by the National Science Foundation Grants PHY-0757058 and PHY-0823459, the University of Birmingham's BlueBEAR HPC service, which provides a High Performance Computing service to the University's research community, the Bondi HPC cluster at the Birmingham Institute for Gravitational Wave Astronomy, as well as resources provided by Supercomputing Wales, funded by STFC grants ST/I006285/1 and ST/V001167/1 supporting the UK Involvement in the Operation of Advanced LIGO.
This research has made use of data or software obtained from the Gravitational Wave Open Science Center (gwosc.org), a service of the LIGO Scientific Collaboration, the Virgo Collaboration, and KAGRA. This material is based upon work supported by NSF's LIGO Laboratory which is a major facility fully funded by the National Science Foundation, as well as the Science and Technology Facilities Council (STFC) of the United Kingdom, the Max-Planck-Society (MPS), and the State of Niedersachsen/Germany for support of the construction of Advanced LIGO and construction and operation of the GEO600 detector. Additional support for Advanced LIGO was provided by the Australian Research Council. Virgo is funded, through the European Gravitational Observatory (EGO), by the French Centre National de Recherche Scientifique (CNRS), the Italian Istituto Nazionale di Fisica Nucleare (INFN) and the Dutch Nikhef, with contributions by institutions from Belgium, Germany, Greece, Hungary, Ireland, Japan, Monaco, Poland, Portugal, Spain. KAGRA is supported by Ministry of Education, Culture, Sports, Science and Technology (MEXT), Japan Society for the Promotion of Science (JSPS) in Japan; National Research Foundation (NRF) and Ministry of Science and ICT (MSIT) in Korea; Academia Sinica (AS) and National Science and Technology Council (NSTC) in Taiwan.
Figures were prepared using \texttt{Matplotlib}~\citep{matplotlib:2007}. Analyses were performed using Bilby~\citep{bilby_paper}, Dynesty~\citep{dynesty}, GWpy~\citep{gwpy}, LALSuite~\citep{lalsuite}, \texttt{Mathematica}~\citep{Mathematica}, NumPy~\citep{numpy:2020}, pyEFPE~\citep{Morras:2025nlp}, and Scipy~\citep{scipy:2020}.

\begin{contribution}
P.S. initiated the research with inputs from G.M. and G.P., and led the writing together with G.P. Bayesian inference on the data and synthetic injections was carried out by G.P., P.S. and G.M. All authors contributed equally to the scientific discussion and interpretation of the results. 
\end{contribution}

\section*{Data Availability}
The strain, PSD and calibration uncertainty data used in this work are openly available via the Gravitational Wave Open Science Center or the LIGO DCC~\citep{GWdata, calibration}. The data behind Fig.~\ref{fig:ecc} are available from Zenodo doi: \href{https://doi.org/10.5281/zenodo.18485978}{10.5281/zenodo.18485978}.

\clearpage
\newpage

\appendix
\section{Additional Parameter Estimation Results}
\label{app:add_pe}

\begin{figure}[h!]
    \centering
    \includegraphics[scale=0.8]{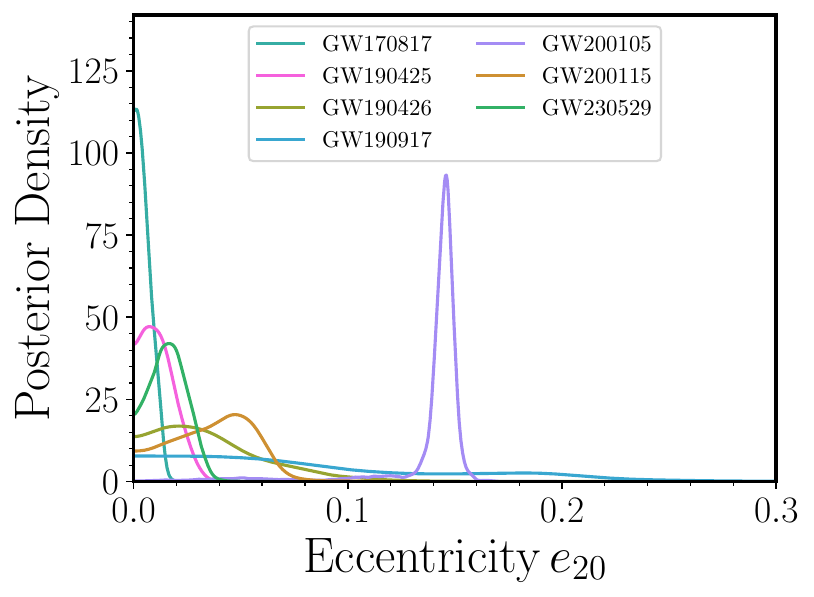}
    \caption{\textbf{Eccentric-precessing analysis of BNS and NSBH events.} One-dimensional posterior distributions for the eccentricity at 20 Hz inferred from Bayesian inference with the eccentric-precessing \EFPE{} model and uniform eccentricity priors on the two BNS events, GW170817 and GW190425, and the NSBH events GW190426, GW190917, GW200115, \gw{} and GW230529. We find that, with the exception of \gw{}, all other analysed events are highly consistent with circular binaries.}
    \label{fig:events}
\end{figure}

In this Appendix, we provide additional parameter estimation results obtained from with \EFPE{} for the analysed BNS and NSBH events. Figure~\ref{fig:events} shows the one-dimensional eccentricity posterior for all seven events, with \gw{} standing out as the only event with a clear measurement of nonzero orbital eccentricity.

\begin{figure*}[h!]
    \centering
    \includegraphics[width=\textwidth]{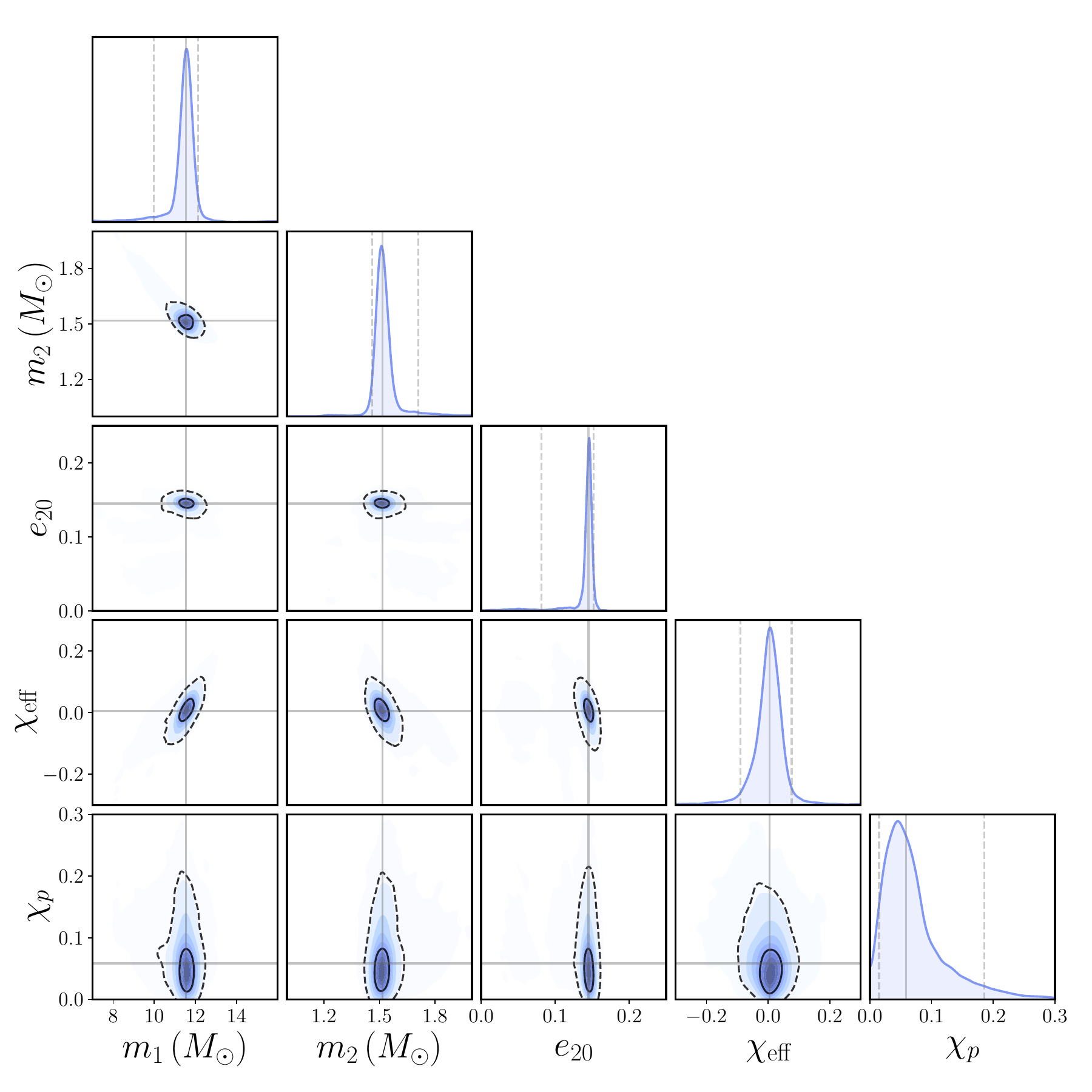}
    \caption{\textbf{Measured properties of \gw{}.} One- and two-dimensional posterior probability distributions of the component masses, effective inspiral spin $\chi_{\rm eff}$, effective precession spin $\chi_p$, and the orbital eccentricity at a GW frequency of $20$ Hz inferred with the eccentric-precessing \EFPE{} waveform model. Solid lines indicate the median value, dashed lines the $90\%$ credible interval. Two-dimensional contours indicate the $50\%$ (solid) and $90\%$ (dashed) credible regions. 
    }
    \label{fig:corner}
\end{figure*}

Focusing on \gw, Fig.~\ref{fig:corner} shows the one- and two-dimensional posteriors for the source-frame masses, the eccentricity, and the two effective spin parameters obtained from our default analysis. We observe the well-known correlation between the masses and $\chi_{\rm eff}$, a very mild correlation between $\chi_{\rm eff}$ and eccentricity, but no correlation between eccentricity and the spin precession parameter $\chi_p$. 

\section{Supporting Investigations \& Robustness Checks}
\label{app:robustness}

\subsection{Waveform Model Uncertainties}
\label{app:waveform_uncertainties}
The eccentric-precessing waveform model \EFPE~only describes the inspiral portion of the signal, and it does not contain higher-order modes (HOMs). We perform numerous tests to determine whether these modelling limitations could falsely induce a non-zero eccentricity posterior.  

\begin{figure}
    \centering
    \includegraphics[width=0.49\textwidth]{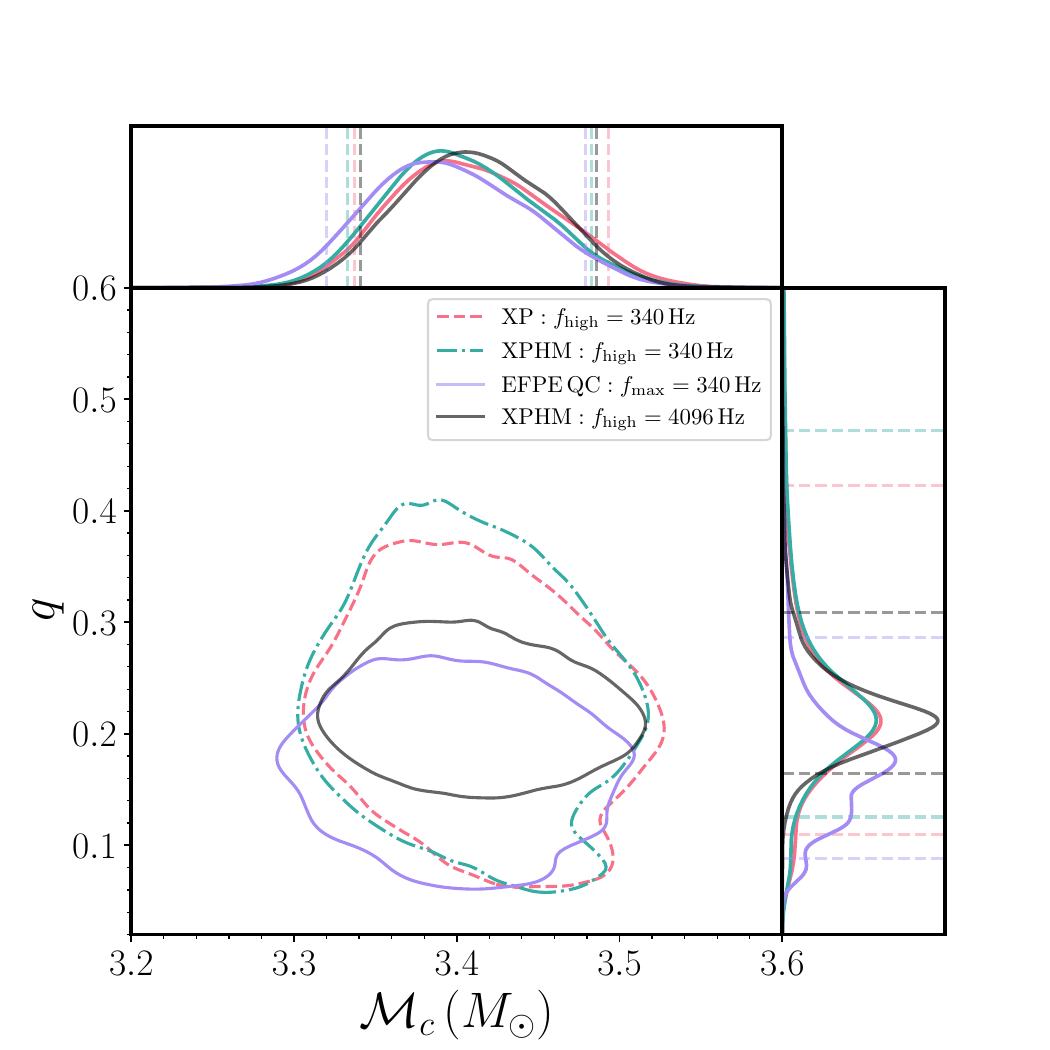}
    \includegraphics[width=0.49\textwidth]{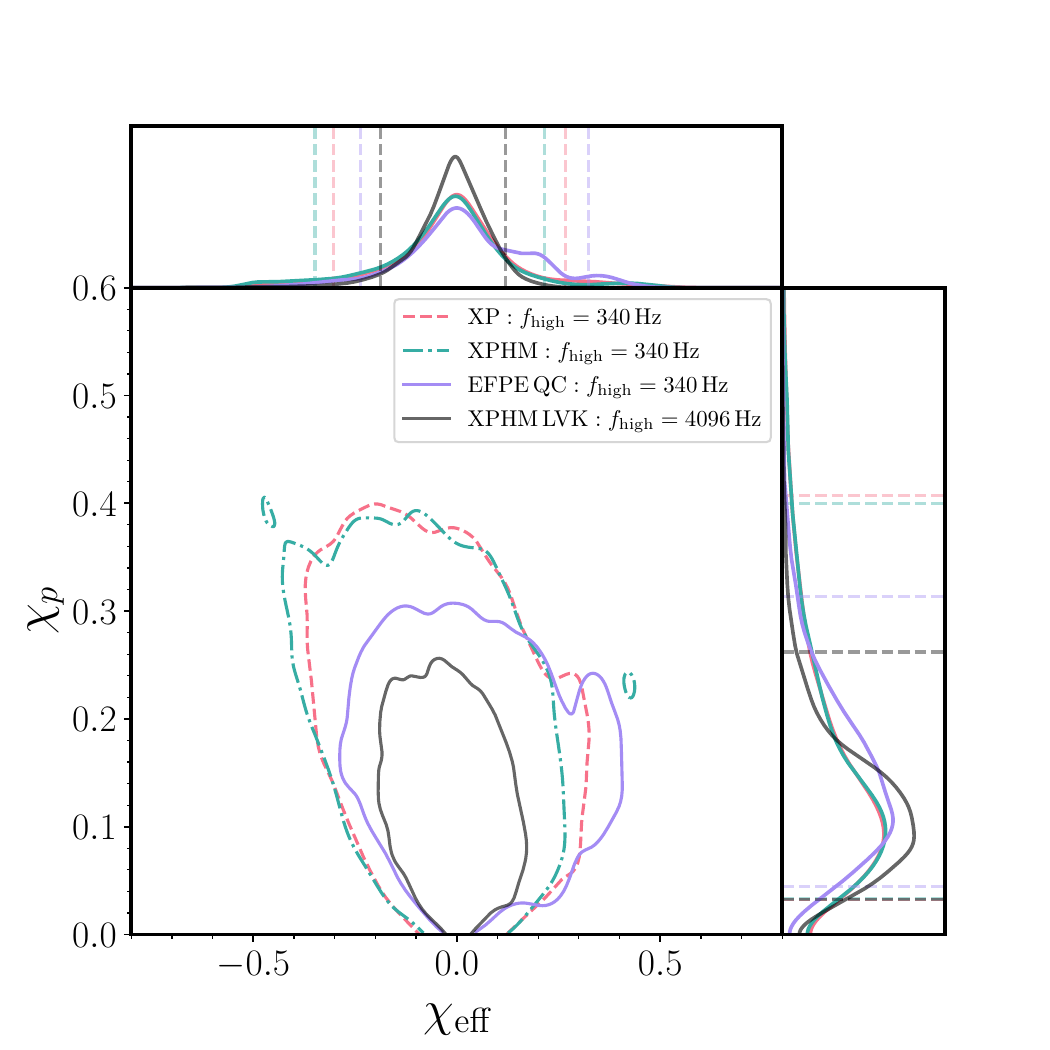}
    \caption{\textbf{Comparison of non-eccentric results of \gw.} One- and two dimensional posterior distributions of the source-frame chirp mass and mass ratio (left) and the effective spin parameters (right) inferred from Bayesian inference with the non-eccentric, precessing waveform models \XP{} and \XPHM{} when imposing a maximum frequency of $340$ Hz in comparison to the \XPHM{} results obtained by the LVK~\citep{GWdata} with $f_{\rm max}=4096$ Hz. While the reduced upper cut-off frequency leads to broader posterior distributions, the results are fully consistent. We also show the results obtained with \EFPE{} restricted to zero eccentricity, which is in good agreement with the \textsc{IMRPhenomX} analyses.}
    \label{fig:LVK-comparison}
\end{figure}

Firstly, we note that the original LVK analyses with non-eccentric waveform models revealed no conclusive support for either HOMs or precession~\citep{LIGOScientific:2021qlt}. Our reanalyses with both \XP~and \XPHM~yield results consistent with those of the LVK even when imposing the restricted $f_{\rm high}$ as opposed to the LVK analyses, which used $f_{\rm high}=4096$ Hz. The one- and two dimensional posteriors for the source-frame chirp mass, the mass ratio and two effective spin parameters are shown in Fig.~\ref{fig:LVK-comparison}. As expected, the restricted frequency range leads to broader but consistent posteriors.

Next, we analyse the event with \EFPE~but restrict the model to zero eccentricity. The inferred posterior distributions are consistent with those obtained from \XP{} and \XPHM{} as also shown in Fig.~\ref{fig:LVK-comparison}, lending further support to a lack of HOMs in \gw~and being suggestive that the merger-ringdown phase adds little additional information for this event. The small differences in the posteriors between \EFPE{} in the quasi-circular limit and the two \textsc{IMRPhenomX} analyses is expected as \EFPE{} does not reduce to those models for zero eccentricity (see Ref.~\cite{Morras:2025nlp} for a detailed discussion).

\begin{figure*}[t!]
    \centering
    \includegraphics[scale=0.8]{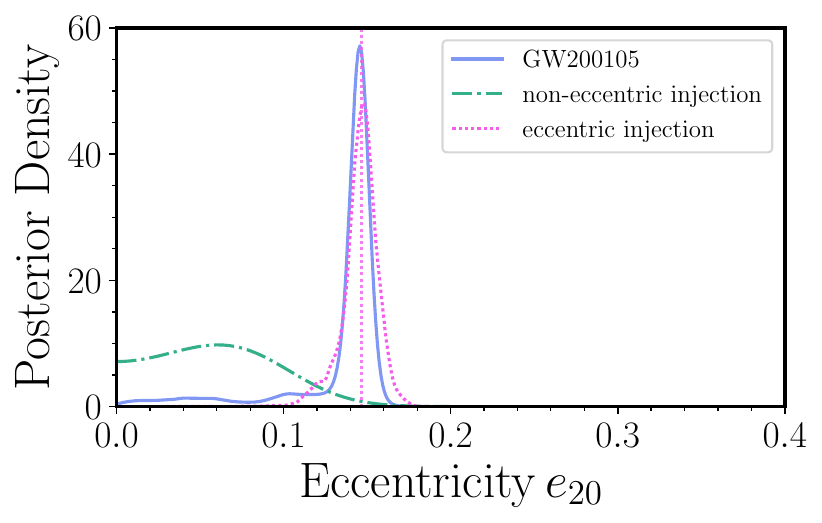}
    \caption{\textbf{Comparison between \gw{} results and injections.} 
    One-dimensional posterior distributions for the eccentricity at $20$ Hz inferred from Bayesian inference with the eccentric-precessing \EFPE{} model on \gw{}, a non-eccentric injection and an eccentric injection into Gaussian noise recoloured by the event PSDs. While small deviations in the parameter recovery are expected as the injections are generated using independent waveform models, we find that \EFPE{} accurately recovers the eccentricity of the eccentric injection (dashed vertical line), and finds a posterior consistent with zero for the non-eccentric injection. }
    \label{fig:injections}
\end{figure*}

We validate the robustness of \EFPE{} in measuring eccentricity by analysing synthetic signals (injections) added to Gaussian noise, recoloured to match the PSDs of \gw{}. 
The setup is otherwise identical to our analysis of \gw{} with \EFPE{} and $f_{\rm high} = 280$ Hz.
We construct two mock signals: one representing a quasi-circular binary, consistent with the masses and spins initially reported by the LVK in~\citep{LIGOScientific:2021qlt}, based on results obtained with \XPHM; and another modelling an eccentric, non-precessing binary with masses, spins, and eccentricity that are consistent with the values reported here using the new \EFPE{} model. 
The non-eccentric mock signal uses \XPHM{} for the injection, while the eccentric mock signal is generated with \textsc{SEOBNRv5EHM}~\citep{Gamboa:2024hli}, an independent non-precessing effective-one-body model that incorporates eccentricity in a fundamentally different way to \EFPE{}.
Both injections include the full inspiral, merger, and ringdown, as well as HOMs.
As shown in Fig.~\ref{fig:injections}, for the non-eccentric injection, we find that the one-dimensional posterior on eccentricity is consistent with zero. 
For the eccentric injection, we accurately recover the injected eccentricity, robustly ruling out zero-eccentricity.
Comparison to the results obtained when analysing \gw{} strongly support the conclusion that this was a low-spin, eccentric NSBH.
If the signal is truly non-eccentric, the absence of HOMs and the merger-ringdown phase in \EFPE{} do not lead to a spurious non-zero measurement of eccentricity. 
Conversely, if the signal is eccentric, \EFPE{} robustly recovers the correct eccentricity, despite differences in the underlying analytical models and how eccentricity is incorporated. 
While this conclusion may not hold across the entire binary parameter space, our tests show no indication that the inference of non-zero eccentricity in \gw{} can be attributed to waveform systematics.

\begin{figure*}[h!]
    \centering
    \includegraphics[width=\textwidth]{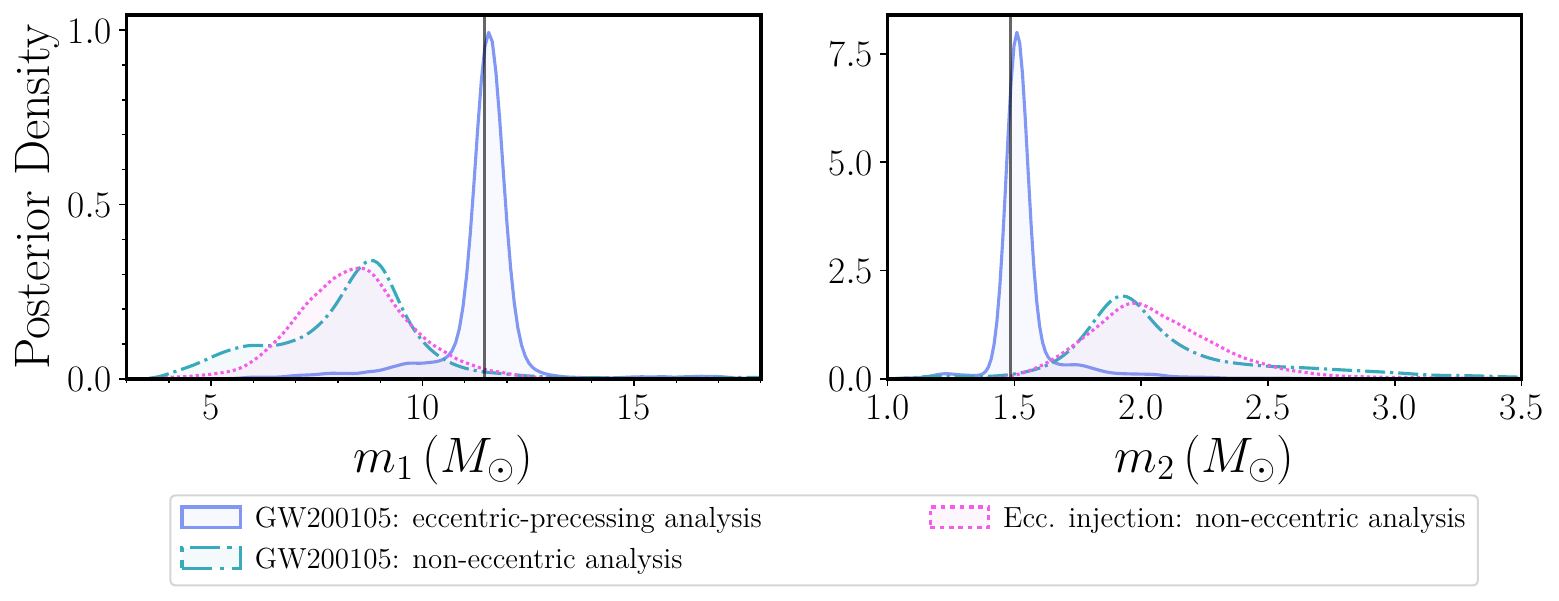}
    \caption{ \textbf{Biases induced by neglecting eccentricity.} 
    One-dimensional posterior distributions for the black hole mass (left panel) and neutron star mass (right panel) from three different analyses.
    The first (blue) uses the eccentric-precessing model (\EFPE{}) to analyse GW200105, the second (green) applies the non-eccentric precessing model (\XPHM{}) to \gw{}, and the third (pink) shows results from a simulated signal based on the best-fit parameters (vertical black line) of GW200105 from the eccentric-precessing analysis, injected into recoloured Gaussian noise.
    Analysing the simulated signal with the non-eccentric, precessing model (pink), we find posteriors that agree with~\citep{LIGOScientific:2021qlt}, showing a clear bias towards a less massive black hole and a heavier neutron star.
    }
    \label{fig:mock-efpe}
\end{figure*}

As a consistency check, we inject a simulated signal into recoloured Gaussian noise using the maximum-likelihood parameters inferred from the eccentric-precessing analysis on GW200105. 
We then analyse this signal with the non-eccentric precessing model \XPHM{} and find that the recovered probability distributions broadly agree with the original inference in~\citep{LIGOScientific:2021qlt}, as shown in Fig.~\ref{fig:mock-efpe}. 
However, when eccentricity is neglected in the analysis, the obtained posteriors demonstrate a strong bias towards a less massive black hole and a heavier neutron star. 

In addition to the \gw{}-specific robustness tests detailed above, we also performed generic validation tests of the \EFPE{} model in~\citep{Morras:2025nlp}, which showed no evidence for a misidentification of eccentricity (see e.g. Fig. 16 therein). Furthermore, we analysed other low-mass BNS and NSBH events with our eccentric-precessing waveform model following the same procedure as for \gw{}. As summarised in Fig.~\ref{fig:events}, we do not find statistically significant evidence for orbital eccentricity in those events~\citep{Morras:2025inprep}, providing further supporting evidence that the clear measurement of orbital eccentricity in \gw{} is robust and not a model artifact.

Finally, we note that the eccentric-precessing model does not include tidal effects. Tidal effects in GWs will typically become important at GW frequencies $f \gtrsim 400$ Hz. Moreover, tides are highly suppressed in unequal mass binaries such as \gw{} and were shown to be negligible in~\citep{LIGOScientific:2021qlt, Gonzalez:2022prs}. 

\subsection{Parameter Estimation Uncertainties}
\label{app:pe_uncertainties}
The results obtained with the eccentric-precessing \EFPE{} waveform model show several interesting features. 
Concretely, we find narrower marginalised PDFs for the individual component masses and mass ratio compared to the non-eccentric analyses. 
To verify the robustness of the fiducial uniform-prior result, we use different sampling methods and sampler settings, such as the number of live points (nlive)~\citep{dynesty}.
To quantify the difference between two one-dimensional probability distributions $p(x)$ and $q(x)$ for a parameter $x$ under changes in the number of live points, we use the Jensen-Shannon (JS) divergence, $D_{\rm JS}$, which is bounded between 0 (identical distributions) and 1 (maximally distinct distributions), and is defined as
\begin{equation}
    D_{\rm JS}(p|q) = \frac{1}{2}[ D_{\rm KL} (p|s) + D_{\rm KL}(q|s)],
\end{equation}
where $s = \frac{1}{2}(p+q)$ and 
\begin{equation}
    D_{\rm KL}(p|q) = \int p(x) \log_2\Bigg(\frac{p(x)}{q(x)}\Bigg) dx,
\end{equation}
is the Kullback-Leibler divergence. To assess the convergence of the \textsc{Dynesty} sampler, we repeat our default analysis which uses ${\rm nlive} = 2000$ also with 3000 and 4000 live points. Relative to the default result, we obtain JS divergences of 0.001 and 0.002, respectively, suggesting that the posterior distributions are statistically indistinguishable. We performed the same check for $f_{\rm high} = 280$ Hz using 1200 and 2000 live point, finding a JS divergence of 0.001.
Additionally, none of the standard \textsc{Dynesty} diagnostics metrics show any evidence for sampling issues and the same maximum likelihood region is consistently found.

\begin{figure}
    \centering
    \includegraphics[scale=0.8]{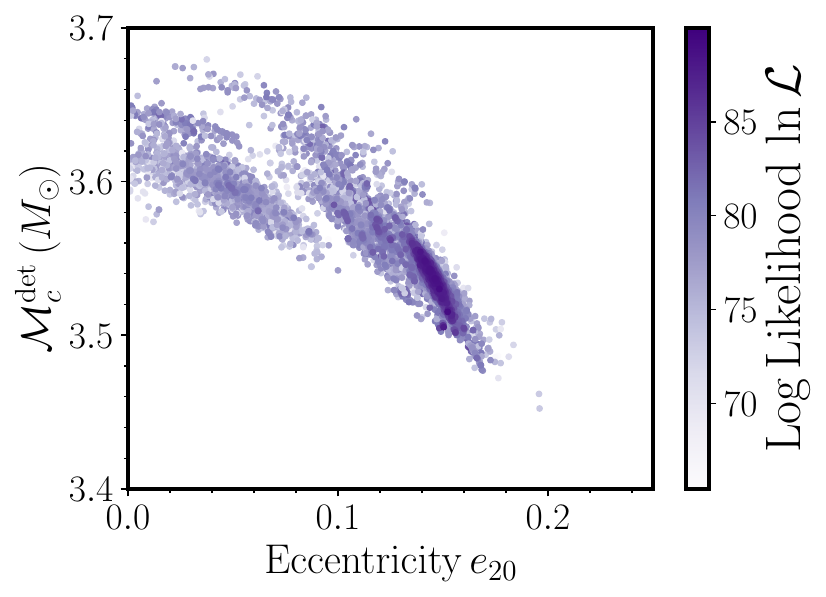}
    \caption{\textbf{Eccentricity--mass correlation.} Two-dimensional distribution of the detector-frame chirp mass and eccentricity posterior samples coloured by their log-likelihood. We observe the expected 0PN mass--eccentricity correlation for low-mass systems. While there is a tail towards smaller values of eccentricity, the bulk of the posterior samples lies in the highest log-likelihood region at $e_{20} \approx 0.15$.}
    \label{fig:LogL_Mc_ecc}
\end{figure}

To assess whether the tail in the eccentricity posterior distribution towards low eccentricities seen in Fig.~\ref{fig:ecc} is physical or driven by sampling artifacts, we show the posterior samples of the detector-frame chirp mass and eccentricity coloured by the log-likelihood in Fig.~\ref{fig:LogL_Mc_ecc}. We observe the well-known 0PN correlation between eccentricity and chirp mass~\citep{Favata:2021vhw} for inspiral-dominated signals. However, the number of samples in the tail is small compared to the highest posterior density region and the log-likelihood is decreased by $\approx$ 10--15
relative to the maximum, which occurs at $e_{20} \approx 0.15$. 

Furthermore, restricting the \EFPE{} model to non-precessing spins, we obtain the same results, ruling out precession as the cause. 
On the other hand, when restricting the model to the leading-order Fourier amplitude mode, we obtain results that are in agreement with \textsc{TaylorF2Ecc}. 
The inclusion of higher-order Fourier harmonics helps break parameter degeneracies, leading to tighter mass parameter constraints.  
These eccentric harmonics are included in the \EFPE{} model but are absent in \textsc{TaylorF2Ecc}.

\begin{figure*}
    \centering
    \includegraphics[scale=0.8]{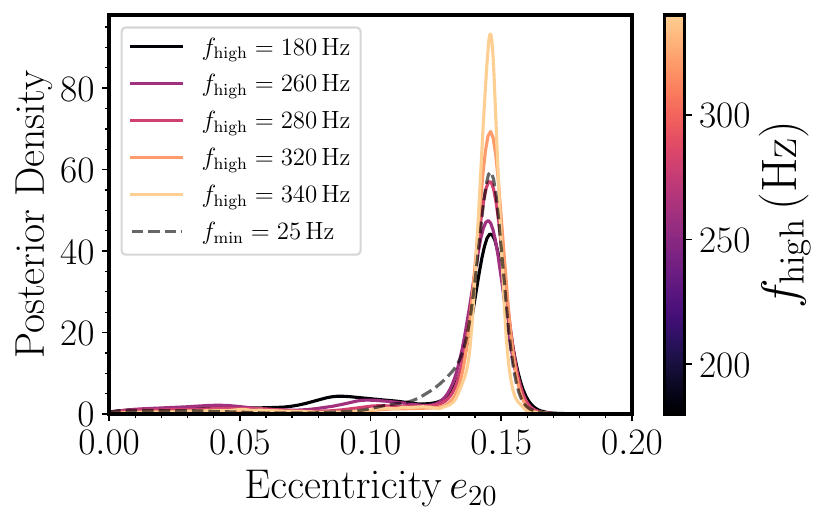}
    \caption{\textbf{Variation of the maximum frequency.} One-dimensional posterior distributions for the eccentricity at $20$ Hz inferred from Bayesian inference with the eccentric-precessing \EFPE{} model on \gw{} using different integration bounds of the likelihood in comparison to our fiducial result for $f_{\rm high}=340$ Hz. We find consistent eccentricity posterior median and mode values even when reducing the frequency range by as much as $160$ Hz, although the posterior distributions broaden and develop a more extended low-eccentricity tail as higher frequency information is removed. The dashed curve shows the result when increasing the lower frequency cutoff from $20$ to $25$ Hz to confirm that the non-zero eccentricity is not induced by residual scattered light noise.}
    \label{fig:fvariation}
\end{figure*}

Another source of uncertainty concerns the high-frequency cutoff of the likelihood integration. Could the early truncation falsely induce a non-zero eccentricity measurement? 
The original LVK non-eccentric analysis integrated the likelihood up to $f_{\rm high} = 4096$ Hz, obtaining a median network SNR of $\sim 13.5 $~\citep{GWdata}.
At our default $f_{\rm high} = 340$ Hz, we find a median SNR of $13.3$ ($13.2$ for $f_{\rm high} = 280$ Hz) for the same waveform model (\XPHM), indicating that we recover the vast majority of the signal despite the reduced bandwidth. Importantly, this is in perfect agreement with expectations for an inspiral-dominated signal, where the majority of SNR is accumulated at low frequencies. In fact, for the non-eccentric analysis, we find that 98\% of the SNR is already accumulated by the time the signal reaches the conservative lower MECO bound of 280 Hz.
In contrast, using the eccentric-precessing \EFPE{} model we find a higher median SNR of $13.7$, providing further indication that the eccentric-precessing model fits the data better. 
As the inspiral-only waveform models cannot be used beyond the inspiral regime, we vary the high-frequency cutoff from $180$ Hz to $340$ Hz to determine whether this truncation impacts the result beyond the recovered signal-to-noise ratio. 
Importantly, we obtain statistically consistent posteriors across a range of different cutoff frequencies, indicating that the eccentricity results presented here are only marginally impacted by our choice of $f_{\rm high}$ as shown in Fig.~\ref{fig:fvariation}. When varying $f_{\rm high}$ between 180 Hz and 340 Hz, the 90\% credible interval shrinks slightly while the median only changes in the third decimal place from $0.143$ for $f_{\rm high}=180$ Hz to $0.145$ for $f_{\rm high}=280$ Hz and higher. The JS divergence between $f_{\rm high}=280$ Hz and $f_{\rm high}=340$ Hz is $0.016$ indicating only very marginal information gain. The observed monotonic increase in log-likelihood and smooth evolution of the posterior with a stable mean are consistent with expectations when more signal is analysed as $f_{\rm high}$ is increased.
We also verify that this behaviour is expected by analysing injections with parameters similar to those measured for \gw{} in zero noise but using the event PSDs. The results we obtain show the same behaviour as seen for \gw{} shown in Fig.~\ref{fig:fvariation}. 
These checks provide strong evidence that the result is robust and not an artifact of the likelihood truncation.

\subsection{Prior Effects}
\label{app:prior_effects}
\begin{figure*}[]
    \centering
    \includegraphics[width=\textwidth]{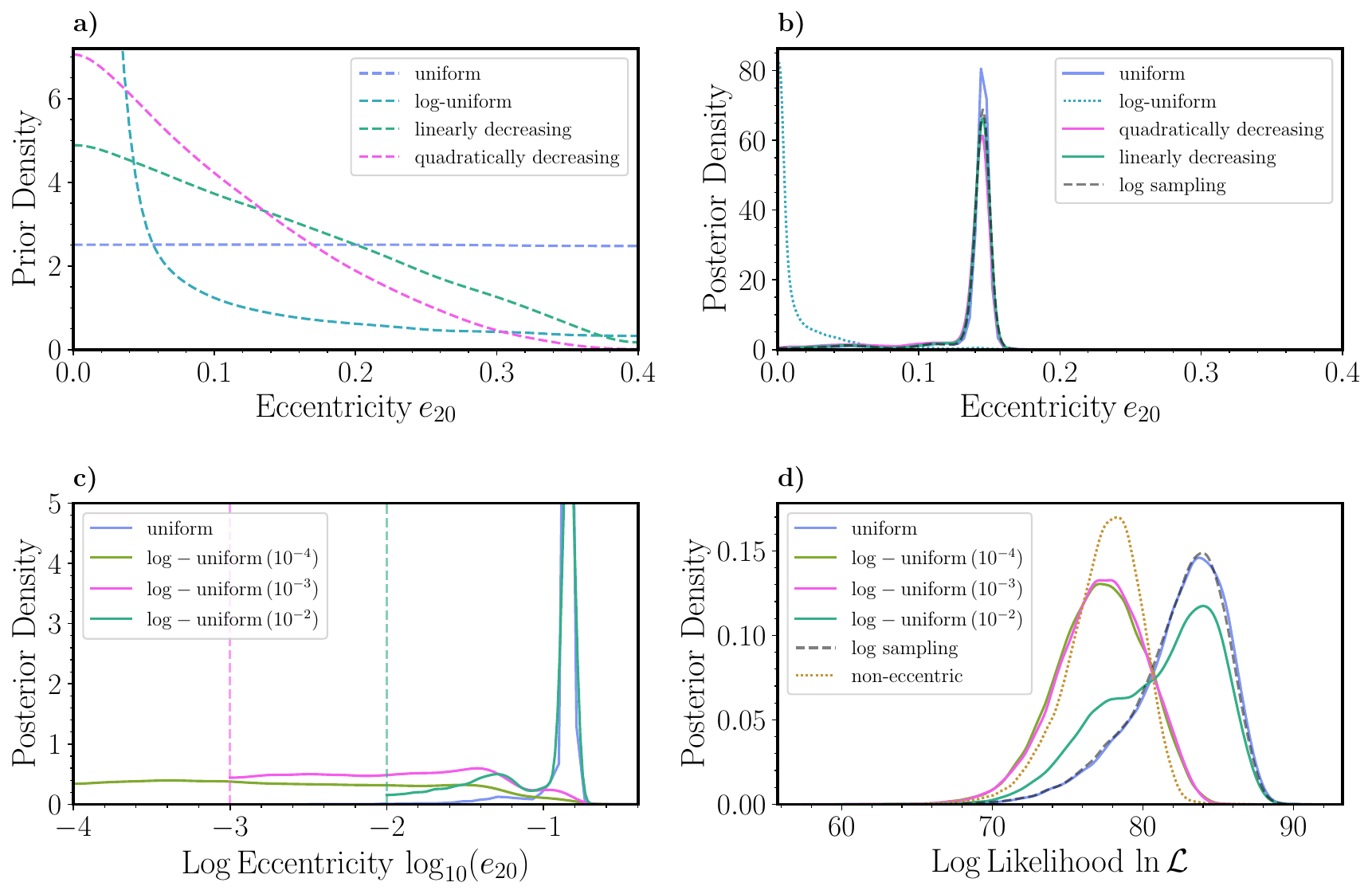}
    \caption{\textbf{Prior dependence.} One-dimensional prior distributions for eccentricity (panel a) and inferred posterior distributions (panel b) for \gw{} under different prior assumptions. Panel b also shows the posterior obtained when sampling in the logarithm of the eccentricity while using the uniform prior (dashed line). Due to the dynamical range of the log-uniform prior, the range of the prior density in panel a) has been restricted for visualisation purposes. Panel c) displays posterior distributions for the logarithm of the eccentricity using the uniform prior and three log-uniform priors with varying lower bounds ($10^{-4}, 10^{-3}$ and $10^{-2}$; indicated by the dashed vertical lines). Only the highly informative log-uniform priors infer zero eccentricity; as the lower bound increases, posterior support shifts away from zero toward the higher log-likelihood mode at non-zero eccentricity identified in all other analyses of \gw{}. This suggests that the posteriors inferred using the log-uniform priors are prior-dominated, residing in regions of lower log-likelihood, as demonstrated in panel d). 
    We caution that the log-uniform results may not be fully converged; see App.~\ref{app:prior_effects} for further details.
    For direct comparison, we also show the log-likelihood for the non-eccentric, precessing waveform model \XPHM{}.}
    \label{fig:priors}
\end{figure*}

In Bayesian inference, the choice of prior can influence the parameter estimates, particularly when the prior is strongly informative relative to the likelihood. 
When using a tightly constraining log-uniform prior for the eccentricity with a lower bound of $10^{-4}$, the resulting posterior is consistent with zero eccentricity.
To verify that the result is an artifact of suppressing the prior volume at eccentricities greater than $\sim 0.1$, we adopt a series of different priors between the log-uniform prior and the agnostic, uniform prior. 
Specifically, we introduce a linearly decreasing prior, $\pi(e) \propto (e_{\rm max} - e)$, and quadratically decreasing prior, $\pi(e) \propto (e_{\rm max} - e)^2$, where $e_{\rm max} = 0.4$. We also consider two additional log-uniform priors with lower bounds of $10^{-3}$ and $10^{-2}$, respectively. 
In addition to changing the prior distribution, we also modified the sampling method for the agnostic uniform prior to sample in the logarithm of the eccentricity. 
This ensures a more comprehensive exploration of small eccentricity values while maintaining equal weights across the entire parameter interval.
The results of these investigations are shown in Fig.~\ref{fig:priors}. 
We find that only the three different log-uniform priors lead to posterior distributions consistent with zero, while all other prior choices exclude zero and identify a consistent median eccentricity of $e_{20}\sim 0.14$. 
We caution that the log-uniform analyses may not be fully converged, as nested sampling converges slowly when the prior volume is concentrated far from the likelihood peak, as is the case here.

Importantly, panel c) demonstrates that increasing the lower log-uniform prior bound from $10^{-4}$ to $10^{-2}$ redistributes posterior probability towards the higher log-likelihood mode that is consistently identified across all other prior (and sampling) choices. 
This is highly suggestive that the posteriors obtained with the log-uniform priors are predominantly driven by the prior and not the data.
Comparing log-likelihoods across different priors (panel d) demonstrates that for low SNR signals, log-uniform priors, despite potential astrophysical motivation, can produce prior-dominated posteriors and lead to inefficient sampling of the high log-likelihood regime. 
This limitation does not affect our fiducial uniform-prior analysis, as the prior and likelihood support overlap and have comparable scales, see Fig.~\ref{fig:priors}b.

For comparison, we also show the log-likelihood of the non-eccentric, precessing analysis with \XPHM{}, which identifies the same low-likelihood mode as found in the analyses with the eccentricity-suppressing log-uniform priors. 
The large log-likelihood difference provides strong evidence that an eccentric signal better characterizes the observed data. 
Finally, we also explored the dependence on the choice of mass priors. 
The default analysis uses priors that are uniform in component masses, even though we sample in $\lbrace\mathcal{M}_c, q \rbrace$. 
The Jacobian when transforming between $\lbrace m_1, m_2 \rbrace$ and the sampling parameters leads to a non-flat prior on $\mathcal{M}_c$ and $q$~\citep{Veitch:2014wba}. 
Using instead a prior that is uniform in $\lbrace\mathcal{M}_c, q \rbrace$ leads to posteriors that are in excellent agreement with the default analysis, implying that the choice of mass priors does not have any impact on the inferred orbital eccentricity. 

\subsection{Noise Artifacts}
\label{app:noise_artifacts}
First, we inject a \EFPE{} mock signal with parameters similar to those obtained from the default analysis into LIGO–Virgo data $84$~s before \gw{} and recover it with the same analysis configuration and event PSD, verifying that the data are normally distributed when whitened with this PSD.
We obtain posteriors that are consistent with the injected values, and the eccentricity is recovered with high fidelity, confidently ruling out a non-eccentric binary. 

Given the significantly higher SNR in LIGO Livingston compared to Virgo, we performed a single-detector analysis using only Livingston data. 
The resulting posteriors are nearly identical to the joint two-detector analysis, confirming that the eccentricity inference is dominated by Livingston. 
This provides a useful check that the result is not driven by potential noise artifacts in Virgo data.

Due to the cleaning procedure applied to the strain data of LIGO Livingston, we perform an additional analysis with \EFPE{} starting from $25$ Hz instead of $20$ Hz, finding excellent consistency between the two results, with $e_{20} = \eccentricityflowtwofive$ in the $25$ Hz analysis as shown in Fig.~\ref{fig:fvariation} (dashed line). We conclude that the applied data cleaning has a negligible impact on the inferred eccentricity in \gw{}. 

\begin{figure*}
    \centering
    \includegraphics[width=0.6\linewidth]{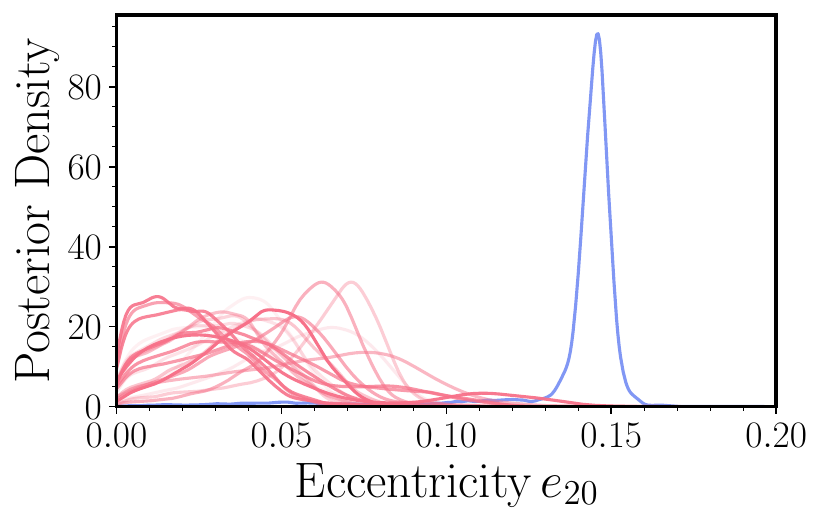}
    \caption{\textbf{Impact of noise fluctuations.} One-dimension posterior distributions (red) of the eccentricity obtained by injection a non-eccentric mock signal into 20 random Gaussian noise realizations weighted by the event PSD and analyzed with \EFPE{}. For comparison, we also show the posterior obtained for \gw{} (blue). While random noise fluctuations can induce deviations from zero, we cannot reproduce a posterior similar to the one for \gw.}
    \label{fig:noise}
\end{figure*}

Finally, to assess whether random noise fluctuations could falsely induce a non-zero eccentricity, we injected the quasi-circular maximum likelihood IMRPhenomXPHM waveform into 20 different realisations of random Gaussian noise recolored by the event PSD and performed Bayesian inference with \EFPE{} on each injection. The one-dimensional eccentricity posteriors are shown in red in Fig.~\ref{fig:noise}. For direct comparison, we also show the posterior inferred for \gw{} from our default analysis. While some noise realisations lead to notable deviations from zero, none of the posteriors is remotely similar to the one obtained for \gw{}.
We then employed a hierarchical empirical Bayes approach~\citep{GelmanBook} to model the population distribution of the upper limit on noise-induced eccentricity. Specifically, we use the 90\% credible upper limit from each noise injection's eccentricity posterior and inferred the hyperparameters of a Beta distribution describing this population using Bayesian inference as implemented in PyMC~\citep{pymc}. This yields a posterior predictive distribution of the form $p(e) \propto e^{15.1} (1 - e)^{209.4}$, representing the variation in noise-driven upper bounds on eccentricity. Using this model, we compute the probability that a noise fluctuation could produce an eccentricity as large as GW200105's median value $(\eccentricitymed)$, finding it to be $2.3 \times 10^{-4}$. This compares the observed median eccentricity of GW200105 to the distribution of noise-induced upper bounds, making this a conservative test, though limited by sample size. The result indicates that such an eccentricity is highly unlikely under the null hypothesis of zero eccentricity in the presence of noise, providing further strong evidence that the observed eccentricity in \gw{} is not due to noise artifacts.

\clearpage

\bibliography{References}{}
\bibliographystyle{aasjournalv7}


\end{document}